\documentstyle[twocolumn,aps]{revtex}

\def\plb{"Phys. Lett. B"}
\def\nub{"Nucl. Phys."}

\begin{document}
\draft
\title{Comment on the broadening of the $^7$Be neutrino line 
in vacuum oscillation solutions of the solar neutrino 
problem\\}
\author{James M. Gelb\cite{gelb}\\}
\address{
Department of Physics\\
University of Texas at Arlington\\
Arlington, Texas 76019\\
}
\author{S. P. Rosen\cite{rosen}\\}
\address{
U.S. Department of Energy\\
Germantown, Maryland 20874\\
}
\date{\today}
\maketitle

\begin{abstract}
We consider the effect of thermal broadening of the $^7$Be neutrino 
on vacuum oscillation solutions of the solar neutrino 
problem and analyze the conditions under which the 
electron-neutrino survival probability must be averaged over 
neutrino energy. For ``just-so" solutions with $\Delta m^2$ of order
$5 \times 10^{-11}~{\rm eV}^2$ averaging is not necessary, 
but for much larger values, it is. We analyze the effective
broadening due to the extended production region in the Sun
and find similar results. We also comment on the possibility
of seasonal variations of the $^7$Be neutrino signal. 
\end{abstract}

\narrowtext

In ``just-so" and other vacuum oscillation solutions of the solar 
neutrino problem\cite{gkk98,justso}, 
it is customary to replace the distance $(L)$ dependent factor 
$P(L,E_\nu) = \sin^2\left({\Delta m^2}L_\odot/E_\nu\right)$ by its
average value of ${1/2}$ when it oscillates 
rapidly over the spectra of pp or $^8$B neutrinos\cite{ssm}.  
The $^7$Be neutrinos\cite{ssm}, however, 
arise from an electron-capture process and are therefore mono-energetic. If 
the thermal broadening of the $^7$Be line\cite{standcross}
is large compared with the energy width of
a single oscillation, then the averaging procedure is still valid; but if the
broadening is small, then averaging is not valid and it becomes necessary to
``fine tune" the oscillation parameters to be consistent with experimental data. 

Here we analyze the conditions under which averaging is or is not valid, and we
show that it is not valid for ``just-so" solutions which appear to be consistent
with the observed anomaly at the
high-energy end of the recoil electron spectrum\cite{gr98}. We also analyze 
the effective broadening induced by the finite size of the solar region in
which $^7$Be neutrinos are produced\cite{standcross}. Finally we study the conditions for
seasonal variations in the $^7$Be neutrino signal\cite{gkk98,many_season}. 

To perform this analysis, we re-express $P(L,E_\nu)$ in terms 
of the energy at which its phase is $\pi/2$\cite{gr98}:
\begin{equation}
1.27~\Delta m^2~{d_\odot\over E_{\pi/2}}~=~{\pi\over 2}~;~~
\Delta m^2 = 8.27\times 10^{-12}~E_{\pi/2}~.	
\end{equation}
For a fixed solar distance $d_\odot$, $1.5\times 10^8$ km, 
this is entirely equivalent
to the usual $\Delta m^2$
characterization with energies in MeV and $\Delta m^2$ in eV$^2$. 
The survival probability of solar
electron-type neutrinos arriving at Earth from the Sun is then given by:
\begin{equation}
P(\nu_e \rightarrow\nu_e;~E_\nu) = 1 - \sin^2 2\theta
\sin^2\left({\pi\over 2}{E_{\pi/2}\over E_\nu}\right)~.
\end{equation}
It reaches its minimum value of $(1- \sin^2 2\theta)$ at an energy of 
$E_{\pi/2}$, and then increases monotonically to 1 as $E_\nu$
increases.

Suppose now that the neutrino energy changes by a small amount:
\begin{equation}
E_\nu \rightarrow E_\nu \pm \delta~.
\end{equation}
Then $P(L,E_\nu)$ becomes:
\begin{equation}
\sin^2\left({\pi\over 2}{E_{\pi/2}\over E_\nu}\right) \rightarrow
\sin^2\left({\pi\over 2}{E_{\pi/2}\over E_\nu}
[1 \mp {\delta\over E_\nu}]\right)~.
\end{equation}
For a significant change in the value of the function,
the additional phase induced by $\pm \delta$ must be of order
${\pi\over 2}$. This will happen when
\begin{equation}
{\delta\over E_\nu} \approx {E_\nu \over E_{\pi/2}}~.
\end{equation}

In the case of $^7$Be neutrinos, the energy and width 
are\cite{standcross}:
\begin{equation}
E_{\nu} = 860~{\rm keV}~;~~\delta \approx 1~{\rm keV}~.
\end{equation}
Fitting the shape of the recoil spectrum requires\cite{gr98}:
\begin{equation}
E_{\pi/2} \approx 6-9~{\rm MeV}~.
\end{equation}
Thus ${\delta/E_\nu}$ is of order $10^{-3}$, while  
${E_\nu/E_{\pi/2}}$ is much larger, namely of order 
$10^{-1}$, and so the condition above is not satisfied. This 
means that the width of the $^7$Be line does not cause more 
than a 1-2$\%$ change in $P(L,E_\nu)$.

We can turn this argument around and ask that, given the 
energy and width of $^7$Be neutrinos, the parameters $E_{\pi/2}$ and
its equivalent $\Delta m^2$ be chosen so as to give a large change
in $P(L,E_\nu)$. From the above analysis, we find that 
$E_{\pi/2}$ must be about one thousand times larger than the 
$^7$Be neutrino energy:
\begin{equation}
E_{\pi/2} \approx 1000~{\rm MeV}~;
\end{equation}
and the corresponding $\Delta m^2$ increases by two orders of
magnitude above the initial case,
\begin{equation}
\Delta m^2 \approx 8 \times 10^{-9}~{\rm eV^2}~.
\end{equation}

Besides the thermal broadening of the $^7$Be line, another feature that
can cause changes in the phase of $P(L,E_\nu)$ is the 
finite extent of the production region in the Sun for $^7$Be neutrinos. 
This region is about $10\%$ of the solar radius\cite{standcross}, 
or $\sim 10^5$ kilometers,
and is small compared with the solar distance $d_\odot$, $\sim
10^8$ km. We can think of it as causing a change in the value of 
$E_{\pi/2}$ by 
one part in a thousand. Such a small fractional change will not have
a significant affect on the distance-dependent factor for $E_{\pi/2}$ in the
range of 6-9 MeV; but for the much larger value of $E_{\pi/2}$ given 
in the last equation it will cause a large variation and it will be 
necessary to average $P(L,E_\nu)$ over the production region.

The variation in the Earth-Sun distance due to the eccentricity of the
orbit of the Earth\cite{gkk98,many_season} 
can also be regarded as a small variation in 
$E_{\pi/2}$. As we have shown before\cite{gr98}, 
this does not cause a significant
seasonal variation in the $^7$Be signal for $E_{\pi/2} = 6~{\rm MeV}$, 
but it does for $E_{\pi/2} = 9~{\rm MeV}$. As $E_{\pi/2}$ grows to about
$50~{\rm Mev}$, the oscillations in the signal become quite 
pronounced\cite{gr98}, 
and for much larger values they eventually become so rapid that again 
an average may need to be performed. The corresponding values of $\Delta m^2$ 
are roughly $8 \times 10^{-11}~{\rm eV}^2$, $4 \times 10^{-10}~{\rm eV}^2$, 
and $4 \times 10^{-9}~{\rm eV}^2$ respectively. 

The need for an average in the eccentricity case is determined by 
the rate at which observations 
of the $^7$Be signal are made. In the SAGE and GALLEX 
experiments\cite{sagegallex}, for example,
the observations are made over approximately 22 days, and so an average will
be required if eccentricity-induced oscillations have a much smaller period. In
experiments like BOREXINO\cite{borexino}, 
the neutrinos will be detected in real time, 
and one might be able to trace out oscillations whose period is long 
compared to the average interval between successive events.

\end{document}